\documentclass[10pt, final, journal, letterpaper, twocolumn]{IEEEtran}

\usepackage{amsfonts}
\usepackage[dvips]{graphicx}
\usepackage{times}
\usepackage{cite}
\usepackage{amsmath}
\usepackage{array}
\usepackage{amssymb}

\newcounter{mytempeqncnt}
\usepackage{stfloats}
\usepackage{slashbox}
\usepackage{graphicx}
\usepackage{footnote}
\usepackage{booktabs}
\usepackage{array}
\usepackage{algorithmic}
\usepackage{algorithm}
\usepackage{threeparttable}
\usepackage{color}

\title{Resource Allocation with Subcarrier Pairing in OFDMA Two-Way Relay Networks}
\author{Hao~Zhang, Yuan~Liu,~\IEEEmembership{Student~Member,~IEEE}, and Meixia~Tao,~\IEEEmembership{Senior~Member,~IEEE}
\thanks{Manuscript received November 16, 2011. The associate editor approving it for publication was Dr. Harish Viswanathan.}
\thanks{The authors are with the Department of Electronic Engineering at Shanghai Jiao Tong University, Shanghai, 200240, P. R. China (e-mail: \{gavinzhanghao, yuanliu, mxtao\}@sjtu.edu.cn).}
\thanks{This work was supported by the NSF of China (60902019) and the Innovation Program of Shanghai Municipal Education Commission (11ZZ19).}

}

\begin{document}
\maketitle
%\makeatletter
\markboth{IEEE Wireless Communications Letters}{}

\vspace{-1.5cm}
\begin{abstract}
This study considers an orthogonal frequency-division
multiple-access (OFDMA)-based multi-user two-way relay network where
multiple mobile stations (MSs) communicate with a common base
station (BS) via multiple relay stations (RSs). We study the joint
optimization problem of subcarrier-pairing based relay-power
allocation, relay selection, and subcarrier assignment. The problem
is formulated as a mixed integer programming problem. By using the
dual method, we propose an efficient algorithm to solve the problem
in an \emph{asymptotically} optimal manner. Simulation results show
that the proposed method can improve system performance
significantly over the conventional methods.
\end{abstract}


\section{Introduction}
\setlength\arraycolsep{2pt}

An important property of orthogonal frequency-division multiplexing
(OFDM)-based relaying is that the frequency diversity can be
exploited by \emph{subcarrier pairing}, which matches the incoming
and outgoing subcarriers at the relay based on channel dynamics and
hence improves system performance. In multi-user environments with orthogonal frequency-division
multiple-access (OFDMA), subcarriers should not only be carefully paired at the relay but
also be assigned adaptively for different users.
If with multiple relays, it further complicates the problem because
relay selection tightly couples with subcarrier pairing and
assignment. Thus, subcarrier-pairing based resource allocation in
multi-user multi-relay OFDMA networks is highly challenging.

Subcarrier-pairing based resource allocation has been originally investigated for single-user single-relay one-way communications (e.g.,
\cite{Hottinen, Hsu}).
In particular, it is proved in \cite{Hottinen} that the\emph{ ordered pairing} is optimal for amplify-and-forward (AF) protocol.
Authors in \cite{Ho} investigated \emph{separated} power allocation
and subcarrier pairing in two-way communication using single relay,
where the power allocation is first employed by water-filling and
then subcarriers are paired at the relay by a heuristic method.
In \cite{Dang}, the subcarrier-pairing based joint optimization of
power allocation, relay selection and subcarrier assignment for
single-user multi-relay systems was studied.
The subcarrier-pairing based joint optimization of power allocation
and subcarrier-user assignment for multi-user single-relay scenario
was studied in \cite{Dong}.
In \cite{YuanTW}, the authors studied relay-assisted bidirectional OFDMA cellular networks, wherein the subcarrier-pairing based joint optimization
of bidirectional transmission mode selection, relay selection, and
subcarrier assignment was investigated by a graph approach. Authors in \cite{YuanTCOM} investigated the jointly optimal
channel and relay assignment for multi-user multi-relay two-way relay networks. These works \cite{YuanTW,YuanTCOM}, however, did not consider power allocation.

In this work, we consider an OFDMA two-way relay network with a common base station (BS), multiple mobile stations (MSs) and multiple relay stations (RSs).
The downlink and uplink traffic for each MS is multiplexed through analog network coding at the RSs.
We formulate a joint optimization problem of subcarrier-pairing based relay-power
allocation, relay selection, and subcarrier assignment. The problem is  a mixed integer programming
problem and we solve it efficiently in dual domain with  polynomial
complexity.

\begin{figure}[tb]
\begin{centering}
\includegraphics[scale=.4]{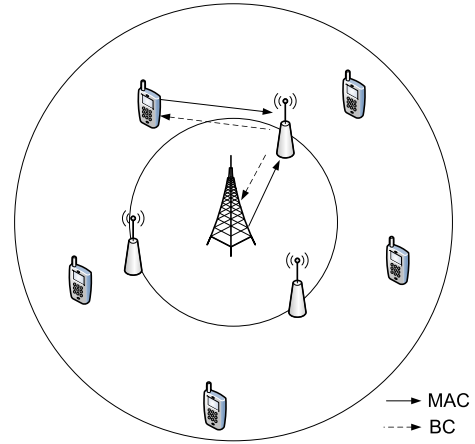}
\vspace{-0.1cm}
 \caption{System model.}\label{fig:system}
\end{centering}
\vspace{-0.3cm}
\end{figure}

\section{System Model and Problem formulation}

We consider a single-cell OFDMA two-way relay
network, as shown in Fig.~\ref{fig:system}, with one BS, multiple
MSs and RSs.
All the MSs are assumed to be cell-edge users so that both the downlink and uplink traffic of each user needs to be relayed through one or more RSs. This assumption is commonly used for cellular relay networks in the literature (e.g., \cite{Dong,Ren,Ng}).
Each RS operates in a half-duplex mode and relays the bi-directional traffic using AF protocol, known as analog network coding.
In specific, the AF two-way relay protocol takes place in two phases \cite{Rankov}. In the first phase, also known as
multiple-access (MAC) phase, all the MSs and the BS concurrently transmit signals while al the RSs listen. In the second
phase, known as broadcast (BC) phase, the RSs amplify the received
signals and then forward them to the MSs and the BS. To avoid multi-user interference, each MS and RS operate in non-overlapping subcarriers in the first and second phases, respectively. The downlink-uplink interference within each user is eliminated by self-interference cancelation.
Furthermore, the channel is assumed to be slowly time-varying and all the channel
state information can be perfectly estimated and known at the BS for
centralized processing.

Let $\mathcal {N}=\{1,2,\cdot\cdot\cdot,N\}$ denote the set of
subcarriers, $\mathcal {K}=\{1,2,\cdot\cdot\cdot,K\}$ denote the set
of RSs, and $\mathcal {U}=\{1,2,\cdot\cdot\cdot,M\}$ denote the set
of MSs.
The channel coefficients from BS $b$ and MS $u$ to
RS $k$ on subcarrier $i$ in the MAC phase are denoted as $h_{b,k,i}$ and $f_{u,k,i}$,
respectively, $\forall u\in\mathcal {U},k\in\mathcal
{K},i\in\mathcal {N}$. In the BC phase, the channel coefficients
from RS $k$ to BS $b$ and MS $u$ on subcarrier $j$ are denoted as
$h_{k,b,j}$ and $f_{k,u,j}$, respectively, $\forall u\in\mathcal
{U},k\in\mathcal {K},j\in\mathcal {N}$. Here channel reciprocity is
used, which is valid in TDD (time-division duplex) mode.
Along with the paths, we further denote $p_{b,k,i}$ and $p_{u,k,i}$
as the transmitted power of BS $b$ and MS $u$ respectively, and $p_{k,u,j}$ as the transmitted power of RS $k$. Then, the sum-rate of uplink-downlink transmission of
MS $u$ over subcarrier pair $(i, j)$ with the assistance of RS $k$
can be expressed as \cite{Rankov,Ho}
\begin{equation}\label{eqn:r1}
\begin{split}
R_{u,k,i,j}=&
\frac{1}{2}\log_2\left(1+\frac{p_{u,k,i}|f_{u,k,i}|^2p_{k,u,j}|h_{k,b,j}|^2}
{p_{k,u,j}|h_{k,b,j}|^2+m_{u,k,i}}\right)\\
+&\frac{1}{2}\log_2\left(1+\frac{p_{b,k,i}|h_{b,k,i}|^2p_{k,u,j}|f_{k,u,j}|^2}
{p_{k,u,j}|f_{k,u,j}|^2+m_{u,k,i}}\right),
\end{split}
\end{equation}
in which
$m_{u,k,i}=1+p_{b,k,i}|h_{b,k,i}|^2+p_{u,k,i}|f_{u,k,i}|^2$.
It can be proved that the sum-rate
$R_{u,k,i,j}$ is concave in the relay power $p_{k,u,j}$.

We then introduce a set of binary variables
$\rho_{u,k,i,j}\in\{0,1\}$ for all $u$, $k$, $i$, $j$. When
$\rho_{u,k,i,j}=1$, it means that subcarrier $i$ in the MAC phase is
paired with subcarrier $j$ in the BC phase and they are used by RS
$k$ to relay the signals of MS $u$. Otherwise, we have $\rho_{u,k,i,j}=0$.
These binary variables must satisfy the following constraints, due to the exclusive subcarrier assignment,
\begin{eqnarray}
\sum_{u=1}^{M}\sum_{k=1}^{K}\sum_{j=1}^{N}\rho_{u,k,i,j}&\leq& 1,
~\forall i, \label{eqn:t1} \\
\sum_{u=1}^{M}\sum_{k=1}^{K}\sum_{i=1}^{N}\rho_{u,k,i,j}&\leq& 1,
~\forall j. \label{eqn:t2}
\end{eqnarray}
For simplicity, we study relay-power allocation and let the transmit
power of the BS and MSs  be fixed.
Each RS is subject to its own peak power constraint. This can be
expressed as:
\begin{eqnarray}\label{eqn:p}
\sum_{u=1}^{M}\sum_{j=1}^{N}p_{k,u,j}\leq P_{k}, ~\forall k,
\end{eqnarray}
where $P_{k}$ is the peak power constraint of RS $k$.

Our objective is to maximize the system total weighted throughput by jointly optimizing the assignment variables $\{\rho_{u,k,i,j} \} $ and the relay power variables $\{ p_{k,u,j}\}$. Mathematically, this can be
formulated as:
\begin{eqnarray}\label{eqn:obj}
&&\max_{\{\boldsymbol{p},\boldsymbol{\rho}\}}\sum_{u=1}^{M}w_u\sum_{k=1}^{K}\sum_{i=1}^{N}\sum_{j=1}^{N}\rho_{u,k,i,j}R_{u,k,i,j}(p_{k,u,j})
\\
&& \textit{s.t.}
~~(\ref{eqn:t1}),(\ref{eqn:t2}),(\ref{eqn:p}),\nonumber
\end{eqnarray}
where $w_u$ is the weight that represents the priority
of MS $u$, $\boldsymbol{p}\in\mathbb{R}_+^{K\times M\times N}$ and
$\boldsymbol{\rho}\in\{0,1\}^{M\times K\times N\times N}$ are
matrices with entries $p_{k,u,j}$ and $\rho_{u,k,i,j}$,
respectively.

\section{Dual Based Algorithm}

We first define $\mathcal {T}$ as the set of all possible
$\boldsymbol \rho$ satisfying (\ref{eqn:t1}) and (\ref{eqn:t2}),
$\mathcal {P}$ as the set of all possible power allocations
$\boldsymbol{p}$ for the given $\boldsymbol{\rho}$ that satisfy
$p_{k,u,j}\geq0$ for $\rho_{u,k,i,j}=1$ and $p_{k,u,j}=0$ for
$\rho_{u,k,i,j}=0$. Denote
$\boldsymbol\lambda=(\lambda_1,\lambda_2,...,\lambda_K)\succeq0$ as
the dual variables associated with the peak power constraints of the
RSs. Then the dual function of the problem in (\ref{eqn:obj}) can be
defined as
\begin{eqnarray}\label{eqn:dual}
g(\boldsymbol\lambda)\triangleq\max_{\begin{subarray}
\boldsymbol{p}\in\mathcal{P}(\boldsymbol \rho)\\ \boldsymbol
\rho\in\mathcal {T}\end{subarray}}L(\boldsymbol p,\boldsymbol
\rho,\boldsymbol\lambda),
\end{eqnarray}
where the Lagrangian is
\begin{eqnarray}\label{Lagrangian}
L(\boldsymbol p,\boldsymbol
\rho,\boldsymbol\lambda)&=&\sum_{u=1}^{M}w_u\sum_{k=1}^K\sum_{i=1}^N\sum_{j=1}^NR_{u,k,i,j}(p_{k,u,j})
\nonumber\\
&&+\sum_{k=1}^{K}\lambda_k\left(P_{k}-\sum_{u=1}^{M}\sum_{j=1}^{N}p_{k,u,j}\right).
\end{eqnarray}

Computing the dual function $g(\boldsymbol\lambda)$ requires us to
determine the optimal $({\boldsymbol{p}, \boldsymbol\rho})$ at the
given dual vector $\boldsymbol\lambda$.
In the following we present the derivations in detail.

\subsection{Optimizing the Primal Variables $(\boldsymbol{p},\boldsymbol{\rho})$ for Given $\boldsymbol{\lambda}$}
We first find the optimal power variables
$\boldsymbol{p}$ by fixing the binary assignment variables
$\boldsymbol{\rho}$. Then we search the optimal $\boldsymbol{\rho}$
by eliminating $\boldsymbol{p}$ in the objective function. Such a
method has been commonly used in the literature (e.g.,
\cite{Hsu,Dang,Dong,Louveaux}).

Let us rewrite
$L(\boldsymbol{p},\boldsymbol\rho,\boldsymbol\lambda)$ as
\begin{equation}
L(\boldsymbol{p},\boldsymbol\rho,\boldsymbol\lambda)=\sum_{k=1}^K\sum_{u=1}^U\sum_{j=1}^NL_{k,u,j}(p_{k,u,j})+\sum_{k=1}^K\lambda_kP_k,
\end{equation}
where
\begin{equation}
\begin{split}
L_{u,k,j}(p_{u,k,j})=w_u\sum_{i=1}^NR_{u,k,i,j}(p_{k,u,j})-\lambda_kp_{k,u,j}.
\end{split}
\end{equation}
Suppose $\rho_{u,k,i,j}=1$ for a certain $(u,k,i,j)$.
It is easy to verify that $L_{u,k,j}(p_{u,k,j})$ is concave in
$p_{k,u,j}$ and thus the optimal $p_{k,u,j}^*(\lambda_k)$ can be
obtained by applying the Karush-Kuhn-Tucker (KKT) conditions. More specifically, $p_{k,u,j}^*(\lambda_k)$ is the
non-negative real root of the following quartic function
\begin{eqnarray}\label{eqn:qua function}
ap_{k,u,j}^4+bp_{k,u,j}^3+cp_{k,u,j}^2+dp_{k,u,j}+e=0,
\end{eqnarray}
where $a,b,c,d,e$ are coefficients determined by the dual variables,
MSs' weights, and channel gains as defined at the top of the next page.

%% =============== Long Equations Here ================
%
\begin{figure*}[!t]
\setcounter{mytempeqncnt}{\value{equation}}
\vspace*{4pt} %\hrulefill
\begin{eqnarray*}\label{coefficients}
a&&=2\ln2\lambda_k|h_{b,k,j}|^4|f_{u,k,j}|^4/m_{u,k,i},\nonumber\\
b&&=4\ln2\lambda_k|h_{b,k,j}|^2|f_{u,k,j}|^2(|f_{u,k,j}|^2+|h_{b,k,j}|^2),\nonumber\\
c&&=2m_{u,k,i}\ln2\lambda_k(|h_{b,k,j}|^4+|f_{u,k,j}|^4+4|h_{b,k,j}|^2|f_{u,k,j}|^2)\nonumber\\
&&-w_u|h_{b,k,j}|^2|f_{u,k,j}|^2(p_{u,k,i}|f_{u,k,i}|^2|f_{u,k,j}|^2+p_{b,k,i}|h_{b,k,i}|^2|h_{b,k,j}|^2),\nonumber\\
d&&=4m_{u,k,i}^2\ln2\lambda_k(|f_{u,k,j}|^2+|h_{b,k,j}|^2)-2w_um_{u,k,i}|h_{b,k,j}|^2|f_{u,k,j}|^2((p_{u,k,i}|f_{u,k,i}|^2+p_{b,k,i}|h_{b,k,i}|^2),\nonumber\\
e&&=2m_{u,k,i}^3\ln2\lambda_k-w_um_{u,k,i}^2(p_{u,k,i}|f_{u,k,i}|^2|h_{b,k,j}|^2+p_{b,k,i}|h_{b,k,i}|^2|f_{u,k,j}|^2).\nonumber\\
\vspace{-0.5cm}
\end{eqnarray*}
\setcounter{equation}{\value{mytempeqncnt}}
\hrulefill
\vspace{-0.3cm}
\end{figure*}
%\vspace{-0.5cm}
%% ============== End ====================

Substituting the optimal power allocations
$\boldsymbol{p}^*(\boldsymbol\lambda)$ into (\ref{eqn:dual}) to
eliminate the power variables, the dual function can be rewritten as
\begin{eqnarray*}%\label{eqn:dual'}
\hspace{-0.2cm}g(\boldsymbol\lambda)=\max_{\boldsymbol \rho\in\mathcal
{T}}\sum_{u=1}^{M}\sum_{k=1}^{K}\sum_{i=1}^{N}\sum_{j=1}^{N}\rho_{u,k,i,j}X_{u,k,i,j}
+ \sum_{k=1}^{K}\lambda_kP_{k},
\end{eqnarray*}
where
\begin{equation}\label{eqn:X}
X_{u,k,i,j}=w_uR_{u,k,i,j}(p_{k,u,j}^*(\lambda_k))-\lambda_kp_{k,u,j}^*(\lambda_k).
\end{equation}

Now we are ready to find the optimal $\boldsymbol{\rho}$.  In the
following, we show that $X_{u,k,i,j}$ defined in (\ref{eqn:X}) plays
an important role in user and relay selection for occupying a
subcarrier pair $(i,j)$.

Noting the constraints (\ref{eqn:t1}) and ({\ref{eqn:t2}}), we
conclude that there is at most one non-zero element for a given
subcarrier pair $(i,j)$. This suggests that at most one MS and one
RS can occupy the subcarrier pair $(i,j)$. Based on the observation,
we define
\begin{equation}\label{eqn:uk}
\mathcal {X}_{i,j}=\max_{k\in\mathcal {K},u\in\mathcal
{U}}X_{u,k,i,j},
\end{equation}
\begin{equation}\label{eqn:uk2}
(u^*, k^*)_{i,j}= \arg\max_{k\in \mathcal {K}, u\in \mathcal {U}}
X_{u,k,i,j}.
\end{equation}
Then the dual function can be further written as
\begin{eqnarray}\label{eqn:simp}
g(\boldsymbol\lambda)=\max_{\boldsymbol \rho\in\mathcal
{T}}\sum_{i=1}^{N}\sum_{j=1}^{N}\rho_{u*^,k^*,i,j}\mathcal {X}_{i,j}
+ \sum_{k=1}^{K}\lambda_kP_{k}.
\end{eqnarray}
From
(\ref{eqn:simp}) it can be seen that if subcarrier $i$ in the MAC
phase is paired with subcarrier $j$ in the BC phase, then the pair
should be used by MS $u^*$ with the help of RS $k^*$, i.e., the MS
and RS with the maximum $X_{u,k,i,j}$ as defined in (\ref{eqn:X}).
This can be interpreted from an economic perspective.
Suppose each dual variable $\lambda_k$ represents the power price of
RS $k$. Then $X_{u,k,i,j}$ can be regarded as the profit of letting
MS $u$ transmitting over the subcarrier pair $(i,j)$ with the help
of RS $k$, where the profit is defined as the throughput revenue
$w_uR_{u,k,i,j}$ minus the power cost $\lambda_k  p^{*}_{k,u,j}$.
Clearly, to maximize the system total profit, each potential
subcarrier pair $(i,j)$ should be assigned to the MS and RS that can
generate the maximum sub-profit.

The remaining problem is then to identify the
optimal subcarrier pairings $\rho_{u^*, k^*, i,j}$. This is  a
standard \emph{two-dimensional assignment problem}. The classical
Hungarian method can be applied to find the optimal
$\boldsymbol{\rho}^*(\boldsymbol\lambda)$ in polynomial time.

\subsection{Optimizing the Dual Vector $\boldsymbol\lambda$}

After computing $g(\boldsymbol\lambda)$, we now solve the standard
dual optimization problem which is
\begin{eqnarray}\label{eqn:g}
&&\min_{\boldsymbol\lambda} ~g(\boldsymbol\lambda)\\
&&s.t.~~\boldsymbol\lambda\succeq0\nonumber.
\end{eqnarray}

Since a dual function is always convex,
subgradient-based methods can be used to minimize
$g(\boldsymbol\lambda)$ with global convergence with the fact that
\begin{equation}\label{eqn:subg}
\Delta\lambda_k=P_{k}-\sum_{u=1}^{M}\sum_{j=1}^{N}p_{k,u,j}^*(\lambda_k)
\end{equation}
is the subgradient at $\lambda_k,\forall k$. In specific, denote
$\Delta\boldsymbol\lambda^{(l)}=(\Delta\lambda_1^{(l)},\Delta\lambda_2^{(l)},...,\Delta\lambda_K^{(l)})$,
then we can update the dual variables as
$\boldsymbol\lambda^{(l+1)}=\boldsymbol\lambda^{(l)}+\omega^{(l)}\Delta\boldsymbol\lambda^{(l)}$. Here, $\omega^{(l)}$ is the diminishing step size at the $l$th iteration to guarantee the convergence of the subgradient
method.

\subsection{Refinement of Power Allocation}
Having the dual point $\boldsymbol\lambda^*$, we now need to
determine the optimal solution to the primal problem
(\ref{eqn:obj}). Due to the non-zero duality gap,  the optimal
$\boldsymbol \rho^*(\boldsymbol\lambda^*)$ and $\boldsymbol
p^*(\boldsymbol\lambda^*)$ may not satisfy all the constraints
(\ref{eqn:t1}), (\ref{eqn:t2}), and (\ref{eqn:p}) in the original
problem. To overcome this problem, we first determine the optimal
assignment $\boldsymbol \rho^*$ in dual domain, and then make a
refinement of the power allocation to meet the power constraints in
the primal problem.
More specifically, denote $\mathcal {A}_{u,k}$ as the set of active
subcarrier pairs assigned to MS $u$ and RS $k$ obtained from the
dual problem. The problem can be written as:
\begin{eqnarray}\label{eqn:refinement}
&&\max_{\boldsymbol{p}}\sum_{u=1}^Mw_u\sum_{k=1}^K\sum_{(i,j)\in\mathcal
{A}_{u,k}} R_{u,k,i,j}(p_{k,u,j})\label{eqn:refp}
\\
&& s.t. ~~\sum_{u=1}^{M}\sum_{(i,j)\in\mathcal
{A}_{u,k}}p_{k,u,j}\leq P_{k}, ~\forall k.
\end{eqnarray}
Clearly, this is a convex problem. By applying KKT conditions, we
can obtain the optimal $p_{k,u,j}^*$ which has the same expression
as that in the dual domain.

Finally we summarize the overall procedure of the
proposed dual-based solution in Algorithm 1. This algorithm is
asymptotically optimal when $N$ is sufficiently large \cite{Yu}.

\begin{algorithm}[tb]
{\small{
\caption{Proposed algorithm for problem (\ref{eqn:obj})}
\begin{algorithmic}[1]
\STATE \textbf{initialize} $\boldsymbol\lambda^{(0)}$ as a random
non-negative vector, $l=0$. \REPEAT  \STATE Compute $X_{u,k,i,j}$
using (\ref{eqn:X}) for all $(u,k,i,j)$ with $p^*_{k,u,j}$ being the
non-negative real root of (\ref{eqn:qua function}).  \STATE Obtain
$\mathcal {X}_{i,j}$ and $(u^*,k^*)$ using (\ref{eqn:uk}) and
(\ref{eqn:uk2}) respectively for all $(i,j)$, then obtain optimal
$\boldsymbol\rho^*(\boldsymbol\lambda^{(l)})$ by solving
(\ref{eqn:simp}). \STATE Update $\boldsymbol\lambda^{(l)}$ using the
subgradients $\Delta\boldsymbol\lambda^{(l)}$ in (\ref{eqn:subg}); Let $l\leftarrow
l+1$. \UNTIL{$\boldsymbol\lambda$ converges.} \STATE Set the final
$\boldsymbol\rho$ as $\boldsymbol\rho^*$ obtained in the dual domain
and refine the power parameter $\boldsymbol p^*$ by solving
(\ref{eqn:refinement}) at the given $\boldsymbol\rho^*$.
\end{algorithmic}
}}
\end{algorithm}

\subsection{Discussion on Complexity and Proportional Fairness}
The complexity of updating the dual variables $\boldsymbol\lambda$
is $\mathcal {O}(K^q)$ (e.g., if the ellipsoid method is used,
$q=2$). The complexity in (\ref{eqn:uk}) and the Hungarian method
are $\mathcal {O}(MK)$ and $\mathcal {O}(N^3)$, respectively.
Combining all, the total complexity of the proposed method is
$\mathcal {O}((MK+N^3)K^{q})$, which is polynomial.

If consider long-term fairness among the MSs, the weight of MS $u$
at time $t$ can be updated by $w_u^{(t)}=1/T_u^{(t)}$, $\forall
u\in\mathcal {U}$, where $T_u^{(t)}$ as the accumulated rate of MS
$u$ at time $t$.
Note that we can let $w_u=1$ for every MS for pure throughput
maximization.

\section{Simulation Results}

We consider a cell with 2 km radius. The RSs are uniformly located
on a circle centered at the BS and with radius of 1 km. The MSs are
randomly but uniformly distributed in the outer circle as in
Fig. 1. The path loss exponent is $4$ and the standard deviation of
log-normal shadowing is $5.8$ dB. The small-scale fading is modeled
by multi-path Rayleigh fading process. A total of $3000$ independent
channel realizations were used. Each channel realization is
associated with a different set of node locations. We set $M=4$,
$K=3$, and $N=32$.
All MS and the BS have the same
maximum power constraints, so do all RSs. We set the BS and MS power
to be $10$ dB per-node and uniformly distributed among all
subcarriers.

As the benchmarks, the performance of Equal Power Assignment (EPA)
based resource allocation and Random Resource Allocation (RRA)
schemes are also presented. Specifically, EPA lets $\boldsymbol p$
be uniformly distributed among all the subcarriers on each relay
station and finds optimal $\boldsymbol \rho^*$ as in Section III-A
proposed algorithm. In RRA, the power is uniformly distributed and
the subcarrier pairs and relays are randomly assigned. The
complexity of the EPA and RRA schemes are $\mathcal {O}(MK+N^3)$ and
$\mathcal {O}(N)$, respectively, which are lower than that of the
proposed algorithm.

\begin{figure}[t]
\begin{centering}
\includegraphics[scale=0.65]{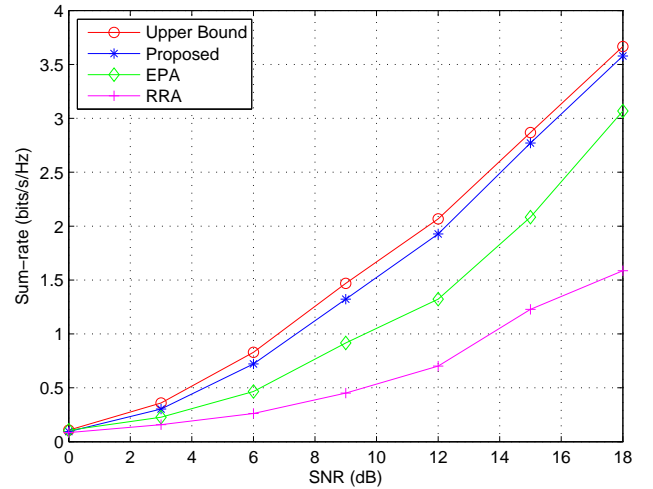}
\vspace{-0.3cm}
 \caption{Average sum-rate versus RS-power per-node.}\label{fig:snr}
\end{centering}
\vspace{-0.3cm}
\end{figure}

Fig.~\ref{fig:snr} compares the average sum-rate achieved by
different schemes.  We first observe that the proposed
dual-based algorithm approaches the upper bound (the optimal dual) closely.
This verifies the effectiveness of the dual method at large
number of subcarriers. One also observes that the proposed algorithm
outperforms the two benchmarks by a significant margin. In
particular,  the proposed algorithm obtains more than $30\%$ and
$200\%$ throughput improvements over the EPA and RRA schemes,
respectively. This tremendous improvement  demonstrates the
superiority of our proposed algorithm.

\section{Conclusion}

In this work, we have studied the subcarrier-pairing based resource
allocation  in OFDMA-based two-way  relay networks. By using the dual
method, an efficient algorithm for joint optimization of
subcarrier-pairing based relay-power allocation, relay selection,
and subcarrier assignment was proposed. Simulation results show that
the proposed algorithm can significantly improve the system
performance compared with the conventional schemes.

\bibliographystyle{IEEEtran}
\bibliography{IEEEabrv,af}

\end{document}